\documentstyle[draft,psfig]{mn}
\newcommand{\mnref}[1]{\hangindent=0.5in \hangafter=1 #1 \par}

\newcommand{\mn}{MNRAS}

\newcommand{\aj}{AJ}
\newcommand{\apj}{ApJ}
\newcommand{\apjs}{ApJS}
\newcommand{\aaa}{A\&A}

\newcommand{\etal}{et al.\ }

\newcommand{\Lsolar}{\mbox{\,$\rm L_{\odot}$}}
\def\gs{\mathrel{\raise1.16pt\hbox{$>$}\kern-7.0pt
\lower3.06pt\hbox{{$\scriptstyle \sim$}}}}
\def\ls{\mathrel{\raise1.16pt\hbox{$<$}\kern-7.0pt
\lower3.06pt\hbox{{$\scriptstyle \sim$}}}}

\title{The infrared luminosity of the torus and the visibility of scattered broad line emission in Seyfert 2 galaxies}
\author[S.L. Lumsden, D.M. Alexander]
{S.L. Lumsden$^{1}$ and D.M. Alexander$^{2}$\\
{}$^1$ {\em Department of Physics and Astronomy,
University of Leeds, Leeds LS2 9JT, UK}\\
{Email -- sll@ast.leeds.ac.uk}\\
{}$^2$ {\em Department of Astronomy \& Astrophysics, 525 Davey Laboratory, Pennsylvania State University, University Park, PA 16802, USA}\\
{Email -- davo@astro.psu.edu}\\
}
\begin{document}

\label{firstpage}

\maketitle

\begin{abstract}
A number of studies have shown that the visibility of scattered broad emission
lines in Seyfert 2 galaxies is strongly dependent on the IRAS $f_{60}/f_{25}$
flux ratio, where those Seyfert 2s with ``warm'' IRAS colours show polarised
broad line emission.  It is now clear that this effect is due to the increasing
dominance of the galactic rather than the AGN emission at 60~$\mu$m in less
luminous ``cool'' Seyfert 2s.  However, we present evidence that the 25~$\mu$m
emission is a good measure of the AGN luminosity for most Seyfert 2s.  Using
this result, we show that the visibility of scattered broad line emission has a
dependence on the AGN luminosity.  The observations can be interpreted
self-consistently if the scale height of the scattering zone varies with
central source luminosity whilst the scale height of the obscuring torus is
approximately constant.
\end{abstract}


\section{Introduction}
The unified model for Seyfert galaxies provides a good qualitative explanation
of the apparently diverse properties of Seyfert 1 and Seyfert 2 galaxies
(e.g.,\ Antonucci 1993).  In this model, both types of Seyfert have the same
nuclear components, namely a broad line region and accretion disc around a
central supermassive black hole, but in Seyfert 2 galaxies the nucleus is
obscured by an optically and geometrically thick dusty torus.  Strong support
for the dusty torus model has come from both theoretical models (e.g. Pier \&
Krolik 1993; Granato \& Danese 1994; Efstathiou \& Rowan-Robinson 1995) and
X-ray and near-IR observations (e.g.,\ Alonso-Herrero, Ward, \& Kotilainen
1997; Turner \etal 1997; Risaliti, Maiolino, \& Salvati 1999). Direct evidence
for hidden Seyfert 1 activity in Seyfert 2 galaxies has come from optical
spectropolarimetric observations which have shown that broad line emission is
observed in scattered flux in some Seyfert 2 galaxies (e.g.,\ Antonucci \&
Miller 1985; Young \etal 1996; Heisler, Lumsden, \& Bailey 1997 hereafter HLB; Moran \etal 2000; Lumsden \etal 2001 hereafter L01; Tran 2001).

HLB carried out the first statistical study of
the frequency of polarised broad line emission in Seyfert 2 galaxies and found
a correlation between the occurrence of polarised broad line emission and the
IRAS $f_{60}/f_{25}$ flux ratio where only those Seyfert 2s with ``warm'' IRAS
colours ($f_{60}/f_{25}<$4.0) showed polarised broad line emission. It was
suggested in that paper that the IRAS colours reflected the the inclination of
the torus, so that ``cool'' sources had more obscured nuclei.  However, this
explanation was later shown to be inconsistent with the evidence provided from
X-ray observations (Alexander 2001; hereafter A01). Instead, A01 proposed an
alternative picture where the IRAS $f_{60}/f_{25}$ flux ratio was related to
the relative luminosities of galactic and AGN activity. In this picture the
lack of polarised broad line emission in ``cool'' Seyfert 2 galaxies is due to
either an intrinsically weak AGN component or a dominant galactic
component. Further analysis by Gu \etal (2001) and L01
supported this picture and it is now clear that the $f_{60}/f_{25}$ flux ratio
is not directly related to the inclination of the torus in the large IRAS
apertures.

However, the previous data suggested that the 25$\mu$m emission was related to
the AGN, and the primary aim of this paper is to test that result by examining
the correlation between the 25~$\mu$m luminosity (hereafter L$_{25}$)
and extinction corrected hard
X-ray (HX, 2--10keV) and [OIII] 5007\AA\ luminosities (hereafter
L$_{\rm HX}$ and L$_{5007}$).  
We use these
results to test how the detectability of polarised broad lines in Seyfert 2s
depends on AGN luminosity, since this has never been properly examined in the
past, and suggest a scheme that explains our results.

\section{Mid infrared emission in Seyfert galaxies}
The most commonly used indicators of the AGN luminosity are 
L$_{\rm HX}$ and L$_{5007}$
(eg.\ Mulchaey et al.\ 1994, Alonso-Herrero et al.\ 1997).
Unfortunately there is no single homogeneous sample of Seyfert galaxies
observed in either band. At the time of writing the most complete compilation
of Seyfert 2 HX properties is presented by Bassani et al.\ (1999) and we have
used this sample for our HX analysis. To provide a comparison, we have included
the Seyfert 1 galaxies from Reynolds (1997).  The heterogeneous nature of these
X-ray samples is not in itself a problem, since the observed galaxies span a
wide range of obscuration and luminosity, which are the primary parameters that
we might expect to introduce a bias in any correlations we find.  However a
potential bias could be introduced by the nature of the observations.  Unlike
the infrared or optical data, the X-ray catalogue is essentially flux limited
(where the limit is set by the available instrument/mission sensitivity).  This
could translate into a false luminosity correlation if there are a large number
of galaxies at the flux limit, but with widely varying redshifts.  We have
attempted to correct for this by only using the galaxies with
$cz<9000$kms$^{-1}$.  We have also excluded all sources from both samples which
are radio-loud or have optical characteristics more typical of LINERs or
quasars so that the sample does at least have a homogeneous optical
classification.  The X-ray emission was corrected for absorption in the case of
Compton thin sources (those with $N_H <1.5\times10^{24}$cm$^{-2}$).  No
correction was made for the Compton thick sources since the direct X-ray
emission is blocked in this case and the detected emission is predominantly
scattered light. Estimating L$_{\rm HX}$ in these sources is dependent on
the scattering model adopted, and hence prone to error; we have not included
these galaxies in the correlation tests involving L$_{\rm HX}$.

The [OIII] 5007\AA\ data was taken from Bassani et al.\ for the Seyfert 2s, and
from Mulchaey et al.\ (1994) for the Seyfert 1s. The [OIII] fluxes given by
Bassani et al.\ are already extinction corrected: the data for the Seyfert 1s
was corrected using, in order of preference, the H$\alpha$/H$\beta$ ratios
given in Mulchaey \etal (1994) and the values of A$_V$ listed in Winkler et
al.\ (1992).  We discarded those galaxies without reported [OIII] fluxes.  IRAS
25$\mu$m fluxes for all these galaxies were taken from, in order of preference,
the IRAS Bright Galaxy Survey (Soifer et al.\ 1989, Sanders et al.\ 1995), the
1.2 Jy catalog (Strauss et al.\ 1990) and the IRAS Faint Source Catalogue
(Moshir et al.\ 1991).  We define L$_{25} = \nu
P_\nu(25\mu{\rm m})$, where $P_\nu$ is the radiant power.  The IRAS data was
not corrected for any residual opacity at this wavelength.  We discarded three
galaxies which either lie in gaps in the IRAS sky coverage (MCG-5-23-16 and
NGC3081) or are blended with other nearby bright sources
(NGC2992 where the IRAS emission is a blend of the emission from
NGC2992 and NGC2993).  We used the survival
analysis techniques in the software package ASURV (version 1.1: La Valley,
Isobe \& Feigelson 1992) to deal with the upper limits present in the IRAS
data.  This implements the methods for correlation analysis outlined in Isobe,
Feigelson \& Nelson (1986).

Figure 1 shows the correlation between L$_{25}$ and (a) L$_{\rm HX}$ and (b)
L$_{5007}$.  Correlations between the three are found at $>99$\% confidence.
We checked for any bias introduced by the essentially flux limited X-ray
samples by also examining the correlations between the fluxes.  Correlations of
the [OIII] 5007\AA\ flux with both the 25$\mu$m and HX fluxes are also found
with $>99$\% confidence.  The correlation of the 25$\mu$m and HX fluxes is
weaker, with only $93$\% confidence.  There is no evidence from this however
for any bias being introduced into the luminosity correlations we have found.

There are trends in the luminosity correlations that the raw probabilities do
not reflect.  The galaxies with lower L$_{\rm HX}$ or L$_{5007}$ tend to have
excess 25$\mu$m emission.  It is likely that there is a significant
contribution from the host galaxy emission (presumably star formation) in these
galaxies.  We tested for the presence of an additional host galaxy component in
two ways.  First, we examined the IRAS colours of the galaxies in the fashion
of Dopita et al.\ (1998).  Figure 2(a) shows that objects with low L$_{5007}$
clearly have IRAS colours typical of starbursts or LINERs
(see also Figure 1 in L01).  Secondly, we examined the compactness of the
10$\mu$m emission as a function of L$_{5007}$, using the
data from Maiolino et al.\ (1995) and Giuricin, Mardirossian and Mezzetti
(1995).  The compactness is defined in the standard sense as the
colour-corrected ratio of the small aperture (3--10$''$) ground based 10$\mu$m
flux and the IRAS 12 $\mu$m flux.  We assume that the 10 and 25$\mu$m emission
trace the same dust (it would be better to measure the compactness at 20$\mu$m
but there are very few small aperture observations available).  Figure 2(b)
shows that there is a strong correlation between the compactness and the AGN
luminosity at $>$99.9\% confidence for galaxies with $cz<5000$kms$^{-1}$.  We
truncated the redshift range in order to avoid false correlations caused by the
fact that the more luminous systems also tend to be more distant.  The reality
of this correlation is demonstrated by the fact that the actual observed
confidence level does not change if we truncate the sample anywhere in the
range $2500<cz<25000$kms$^{-1}$.  There is clear evidence from both these
indicators that the host galaxy can be a significant contributor at these
wavelengths in less luminous Seyferts.

We estimated the luminosity at which galactic 25$\mu$m emission has a
significant effect on the overall spectral energy distribution by
inspection of Figure
2(a).  We found that galaxies with L$_{\rm HX}\ls10^{7.2}$\Lsolar\ and
L$_{5007}\ls10^{7.3}$\Lsolar\ were likely to have IRAS colours more typical of
the host galaxy.  We therefore excluded these systems when deriving the fits to
the data shown in Figure 1.  Note however that the actual significance of the
correlations seen in Figure 1 hardly changes if we exclude these lower
luminosity AGN.  We fitted the data with the bisector of the linear regression
of $y$ on $x$ and $x$ on $y$, which is the best estimate of the underlying
trend in this case (see Isobe et al.\ 1990) The slopes found are all close to
unity, within the errors, as expected if all three observational parameters
reflect the underlying AGN luminosity.  The actual values are 0.82$\pm$0.12 and
1.05$\pm$0.12 for the data shown in Figure 1(a) and (b) respectively, and
0.98$\pm$0.08 for the correlation between L$_{\rm HX}$ and L$_{5007}$
(not shown in Figure 1, since it essentially repeats the previous work of
Mulchaey et al.\ 1994, and Alonso-Herrero et al.\ 1997).  If we include the
galaxies with excess emission at 25$\mu$m the slope decreases to 0.72$\pm0.06$
in both Figure 1(a) and (b).  This is in the sense expected if the lower
luminosity galaxies do have excess 25$\mu$m emission. It is clear from these
results that the L$_{25}$ is as good an indicator of the underlying AGN
luminosity as the extinction corrected X-ray or [OIII] luminosities, at least
for those galaxies with $L_{5007}>10^{7.3}$\Lsolar.



\section{Detectability of scattered broad line emission in Seyfert 2 galaxies}
Scattered broad H$\alpha$ from the broad line region has been detected in $>30$
Seyfert 2s using optical spectropolarimetry since the initial observations of
NGC1068 by Antonucci \& Miller (1985). Until the work of HLB there were no
statistically significant samples to test the mechanisms that effect the
visibility of scattered broad line emission in Seyfert 2 galaxies.  Now with
the addition of the larger distance limited sample of Moran \etal (2000), the
far infrared flux and luminosity limited sample of L01 and the heterogeneous
optical and mid infrared selected sample of Tran (2001) we have a much firmer
grasp of how often we see such {\em hidden} broad line regions (HBLRs).
However, there have been no direct attempts to relate their detectability
directly to the AGN luminosity, partly because suitable measures of that
luminosity were not always available.  Therefore, we have considered this issue
using L$_{25}$ as an indicator of the AGN luminosity using the above samples;
we have implicitly assumed that all of the Seyfert 2s in the Moran \etal (2000)
sample are not HBLRs unless they are included in their reported detections or
were previously known to contain HBLRs (though we have included the two
extra detections given in Moran et al.\ 2001).  We have used the IRAS fluxes
from the same sources as in Section 2.  Some of the Moran et al.\ sample lie
within gaps in the IRAS sky coverage and were excluded (MCG-5-23-16, NGC3081,
NGC4117 and NGC7450), as was NGC2992 for the reason outlined in Section 2.
Finally, we note that some of the galaxies in the Moran et al.\ sample only
have upper limits at 25$\mu$m, necessitating the use of the ASURV package 
when comparing samples on the basis of luminosity.

We tested whether the redshift distribution of the HBLRs and non-HBLRs are the
same for all three samples before comparing the luminosities to avoid
introducing a Malmquist type bias into our analysis.  A Mann-Whitney U test
shows there is no evidence for a difference in the redshift distributions
of HBLRs and non-HBLRs in the Moran et al.\ (2000) and
L01 samples at the 5\% significance level.  The Tran (2001)
sample does however show a difference at the 1\% significance level.  This is
largely due to the presence of many low luminosity nearby AGN in the 12$\mu$m
galaxy sample of Rush et al.\ (1993) which makes up well over 50\% of the Tran
sample.  If we impose a lower redshift limit on the Tran sample of
2500kms$^{-1}$, then the difference in the redshift distribution is removed.
Therefore in what follows we have used this truncated version of the Tran
sample which has 2
HBLRs and 8 non-HBLRs removed.  In practice, this has little
effect on the actual significance of the difference in luminosities between the
subsets found below, but is the simplest way to correct for this bias in the
Tran sample.  We also note that we have excluded the 2 galaxies from the
L01 sample which had insufficient signal-to-noise to adequately
determine the absence of an HBLR (see Section 3.1 of L01 for
further details).

Figure 3 (a-c) shows the L$_{25}$ distributions for the samples.  
Statistically the distributions for
the HBLR samples differ from the non-HBLR samples at the 99
level for both the Moran et al.\ and Tran samples (Figure 3a and 3b),
with the galaxies containing HBLRs being more luminous.  We used the four
available two sample tests in ASURV to test whether the upper limits present
affected the result: the statistical significance was $>99$\% for all
four.  These tests are 
outlined in Feigelson \& Nelson (1985).  There are fewer low luminosity
Seyferts in the L01 sample, so the clear distinction present in the other
samples is less obvious in Figure 3(c).  Statistically the galaxies with and
without HBLRs do have different L$_{25}$ at the 88\% confidence level, but this
cannot be described as conclusive.  We therefore also considered L$_{5007}$,
which was available for this sample.  The two samples differ here at the 92\%
confidence level in good agreement with the results of the 25$\mu$m data.
Again this is not conclusive, but the similarity of the two indicators gives
added confidence to our results for the Moran et al.\ and Tran samples, and the
actual result agrees with the trends seen in those samples.

Finally we considered how our results are affected by our caveat that L$_{25}$
may not truly reflect the AGN luminosity in the lowest
luminosity galaxies.  Clearly, all this can do is move some of the low
luminosity galaxies (which uniformly show no evidence of an HBLR) to higher
luminosities.  Therefore the actual distinction between galaxies containing an
HBLR and those without may be greater than we have estimated.  

\section{Discussion and conclusions}
We have shown that the large beam 25$\mu$m emission measured by IRAS 
is a good indicator of the underlying
AGN luminosity in Seyfert galaxies for even the most obscured systems.
Within the Unified model of Seyfert galaxies, any AGN
emission at 25$\mu$m is produced by dust emission within the torus.  Our
results therefore suggest a direct relationship between the luminosity of the
central source and the luminosity of the dusty torus, as was previously found
for shorter wavelength infrared data by Alonso-Herrero et al.\ (1997).
Furthermore, those sources with powerful AGN have warm IRAS $f_{60}/f_{25}$
colours.  A corollary of this and previous results of HLB
is that those sources with luminous 25$\mu$m emission should therefore be more
likely to have detectable HBLRs; we have directly shown this is the case for 3
independent optical spectropolarimetric samples.

An obvious conclusion for the lack of a HBLR could be that they do not exist in
the lower luminosity Seyfert 2 galaxies (ie these are ``true'' Seyfert 2s).
L01 discussed this issue, where it was shown that there is no significant
difference in the detection rates of obscured HX emission or compact cm radio
emission in galaxies with and without HBLRs.  A similar result is shown
directly by the HX study of Moran et al.\ (2001) and can also be inferred for
the Moran et al.\ (2000) sample from the analysis of Gu et al.\ (2001).
Clearly both galaxies with and without HBLRs show evidence for obscured or
scattered Seyfert 1 cores in HX spectra.

How can the luminosity of the AGN affect the ``visibility'' of the HBLR?
One possibility, addressed in detail in A01 and L01,
is that the relatively larger contribution from
the host galaxy in the less luminous AGN leads to increasing
dilution of the polarisation signal, making it harder to detect
any HBLR.  Alternatively, we might be seeing a more direct physical connection
between the AGN luminosity and the visibility of the HBLR.
The scale height of the scattering particles will be larger in the more
luminous sources for any reasonable geometry and electron density
distribution. This creates two biases against the detection of HBLRs in less
luminous sources. First, any HBLR emission will be intrinsically weaker (e.g.,\
see A01) and secondly, the angular size of the scattering screens
will be smaller.  A smaller scattering screen has a larger probability of being
obscured from view by either dust in the host galaxy or dust within the torus.
The importance of the latter effect will depend upon the scale height of the
torus and the relationship this has with the luminosity of the central
source. There are several processes that could effect the scale height of the
torus (e.g.,\ radiation pressure will inflate the torus, gravitational and
dynamical effects will tend to compress it).  However, there are no forces that
will tend to compress the size of the scattering volume.  Therefore, the
dependence on luminosity of the scale height of the scattering particles should
be greater than that of the torus.  The fraction of scattering that is obscured
by the torus will therefore decrease as the AGN luminosity increases, naturally
giving rise to the observed behaviour found in Section 3.

\section*{Acknowledgements} 
SLL acknowledges support from PPARC through the award of an Advanced Research
Fellowship.  DMA acknowledges financial support from the NSF CAREER award
AST-9983783 (P.I. W.N.Brandt).  We thank the referee Ed Moran for his very
helpful comments.

\parindent=0pt

\vspace*{-8mm}

\section*{References}
\mnref{Alexander, D.M., 2001, \mn, 320, L15 (A01)}
\mnref{Alonso-Herrero, A., Ward, M.J., Kotilainen, J.K., 1997, \mn, 288, 977}
\mnref{Antonucci, R., Miller, J.S. 1985, ApJ, 297, 621}
\mnref{Antonucci, R. 1993, ARA\&A, 31, 473}
\mnref{Bassani, L., Dadina, M., Maiolino, R., Salvati, M., Risaliti, G.,
        della Ceca, R., Matt, G., Zamorani, G., 1999, \apjs,  121, 473}
\mnref{Dopita, M.A., Heisler, C.A., Lumsden, S.L., Bailey, J.A., 1998, \apj, 
	498, 570}
\mnref{Efstathiou, A., Rowan-Robinson, M.\ 1995, \mn, 273, 649}
\mnref{Feigelson, E.D., Nelson, P.I., 1985, \apj,  293, 192}
\mnref{Giuricin, G., Mardirossian, F., Mezzetti, M.\ 1995, \apj, 438, 527}
\mnref{Granato, G.L., Danese, L.\ 1994, \mn, 268, 235}
\mnref{Gu, Q., Maiolino, R., Dultzin-Hacyan, D., 2001, \aaa, 366, 765}
\mnref{Heisler, C.A., Lumsden, S.L., Bailey, J.A., 1997, Nature, 385, 700 (HLB)}
\mnref{Isobe, T., Feigelson, E.D., Nelson, P.I., 1986, \apj,  306, 
        490} 
\mnref{Isobe, T., Feigelson, E.D., Akritas, M.G., Babu, G.J., 1990,
	\apj, 364, 104}
\mnref{LaValley, M., Isobe, T., Feigelson, E.D., 1992, BAAS, 24, 839}
\mnref{Lumsden, S.L., Heisler, C.A., Bailey, J.A., Hough, J.H., Young, S.,
	2001, \mn, in press (L01)}
\mnref{Maiolino, R., Ruiz, M., Rieke, G.H., Keller, L.D., 1995,
        \apj,  446, 561}
\mnref{Moran, E.C., Barth, A.J., Kay, L.E., Filippenko, A.V., 2000, \apj,
        540, L73}
\mnref{Moran, E.C., Kay, L.E., Davis, M., Filippenko, A.V., 
	Barth, A.J., 2001, \apj, 556, L75}
\mnref{Moshir, M., et al., 1991, Explanatory Supplement to the IRAS Faint
        Source Survey, Version 2.  JPL, Pasadena}
\mnref{Mulchaey, J.S., Koratkar, A., Ward, M.J., Wilson, A.S., Whittle, M.,
	Antonucci, R.R.J., Kinney, A.L., Hurt, T., 1994, \apj, 436, 586}
\mnref{Pier, E., Krolik, J.\ 1993, \apj, 418, 673}

\mnref{Reynolds, C.S., 1997, \mn 286, 513}
\mnref{Risaliti, G., Maiolino, R., Salvati, M.\ 1999, \apj, 522, 157}
\mnref{Rush, B., Malkan, M.A., Spinoglio, L.\ 1993, \apjs, 89, 1}
\mnref{Sanders, D.B., Egami, E., Lipari, S., Mirabel, I.F., Soifer, B.T.,
        1995, \aj, 110, 1993}
\mnref{Soifer, B.T., Boehmer, L., Neugebauer, G., Sanders, D.B., 1989,
        \aj, 98, 766}
\mnref{Strauss, M.A., Davis, M., Yahil, A., Huchra, J.P., 1990, \apj, 361, 49}
\mnref{Tran, H., 2001, \apj, 554, L19}
\mnref{Turner, T.J, George, I.M., Nandra, K., Mushotzky, R.F., 1997, ApJ,
        488, 164}
\mnref{Winkler, H., Glass, I.S., van Wyk, F., Marang, F., Jones, J.H.S.,
	Buckley, D.A.H., Sekiguchi, K., 1992, \mn, 257, 659}
\mnref{Young, S., Hough, J.H., Efstathiou, A., Wills, B.J., Bailey, J.A., Ward,
M.J., Axon, D.J.\ 1996, \mn, 281, 1206}

\newpage

\onecolumn

\vspace*{5mm}

\begin{center}
\begin{minipage}{6.5in}{(a)\hspace*{5mm}\vspace*{-15mm}\hspace*{2.8in}(b)
\psfig{file=Figure1.ps,width=6.5in,angle=0,clip=}
}\end{minipage}
\vspace*{10mm}

\begin{minipage}{\textwidth}{
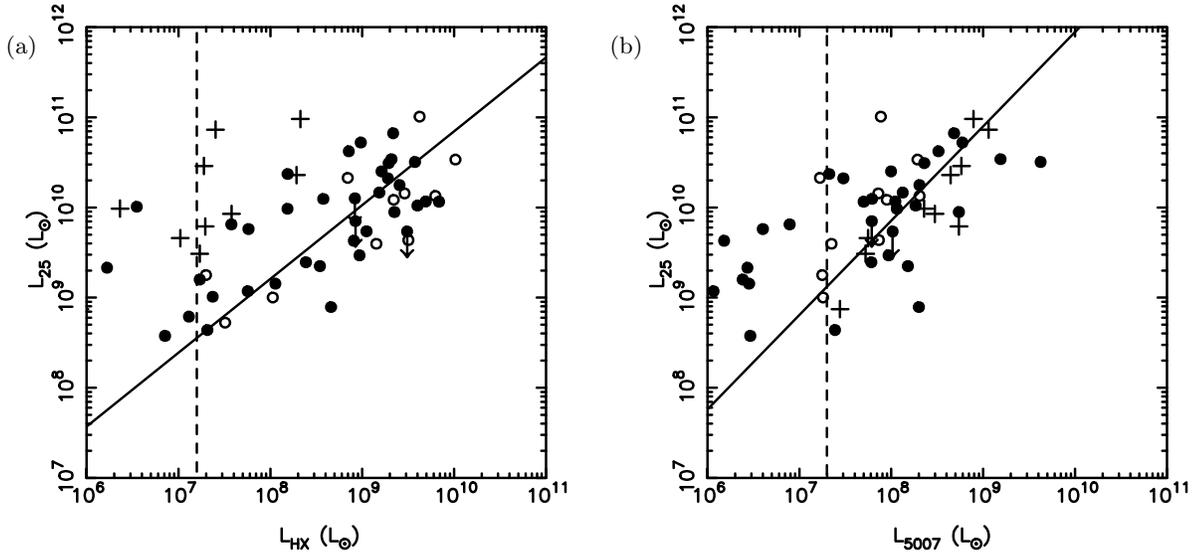
{\bf Figure 1:}  The observed correlations between L$_{25}$ and (a)
L$_{\rm HX}$ and (b)
L$_{5007}$.  The Seyfert 1s are shown as {\tt o}, the
Compton thin Seyfert 2s as $\bullet$ and the Compton thick Seyfert 2s as
$+$.  25$\mu$m upper limits are shown by the arrows.
The solid straight line in each plot is the best fit to the data
excluding those galaxies which are Compton thick, and
those galaxies in (a) and (b)
L$_{\rm HX}\ls10^{7.2}$\Lsolar\ and
L$_{5007}\ls10^{7.3}$\Lsolar.
The vertical dashed lines show the boundaries between those galaxies
which we have classed as AGN dominated, and those in which the host galaxy
dominates.  Excess 25$\mu$m emission can clearly be seen for the galaxies
to the left of these lines.
}\end{minipage}
\end{center}

\vspace*{5mm}

\begin{center}
\begin{minipage}{6.5in}{(a)\hspace*{5mm}\vspace*{-15mm}\hspace*{2.8in}(b)
\psfig{file=Figure2.ps,width=6.5in,angle=0,clip=}
}\end{minipage}
\vspace*{7mm}

\begin{minipage}{\textwidth}{
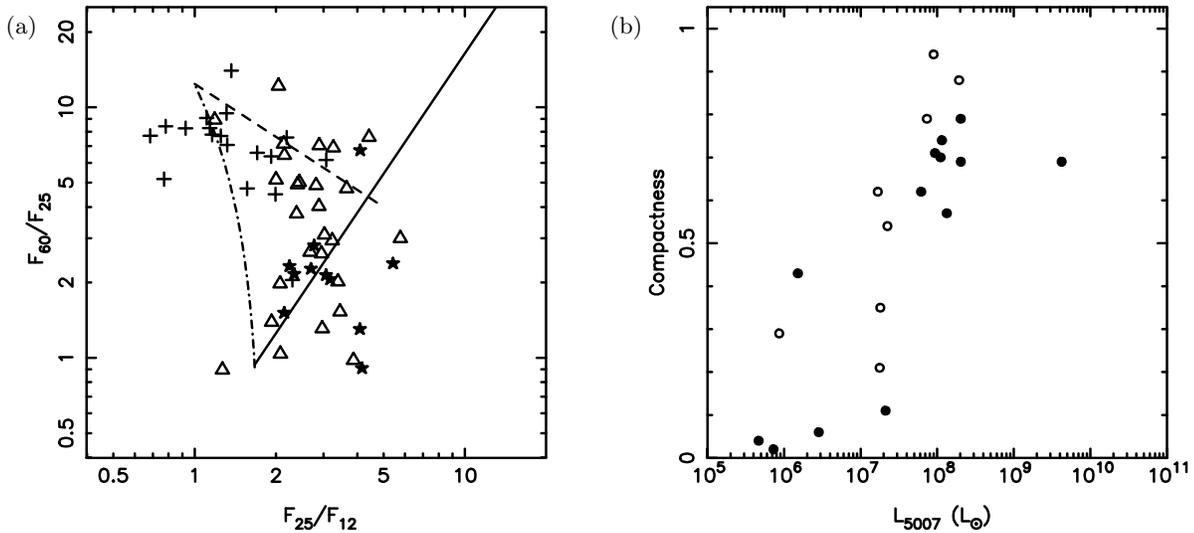
{\bf Figure 2:} (a) IRAS colours of the sample as a function of L$_{5007}$.
The Seyfert 1s and 2s have been merged in this sample.  Those with
L$_{5007}<10^{7.3}$\Lsolar\ are shown as +.  These are the galaxies which we
class as host galaxy dominated.  Those with
$10^{7.3}<$L$_{5007}<10^{8.5}$\Lsolar\ are shown as $\triangle$.  Those with
L$_{5007}>10^{8.5}$\Lsolar\ are shown as $\star$.  The solid line plotted is
the track followed by an increasingly reddened Seyfert 1.  The dashed line
shows the locus of the starburst/LINER population.  The dot-dashed line is the
boundary within which galaxies of mixed AGN and starburst excitation should
lie.  See Dopita et al.\ (1998) for more details.  (b) The compactness of the
10$\mu$m emission as a function of the L$_{5007}$.  Seyfert 1s are shown as
{\tt o}, and Seyfert 2s as $\bullet$, clearly indicating the lower luminosity
objects are more likely to be less compact. Only galaxies with
$cz<5000$kms$^{-1}$ are shown. }\end{minipage}
\end{center}

\vspace*{5mm}

\begin{center}
\begin{minipage}{3.25in}{(a)\hspace*{5mm}\vspace*{-1.7in}
\psfig{file=Figure3a.ps,width=3in,angle=0,clip=,bbllx=0pt,bblly=0pt,bburx=272pt,bbury=418pt}}\end{minipage}
\begin{minipage}{3.25in}{(b)\hspace*{5mm}\vspace*{-1.7in}
\psfig{file=Figure3b.ps,width=3in,angle=0,clip=,bbllx=0pt,bblly=0pt,bburx=272pt,bbury=418pt}}\end{minipage}
\vspace*{7mm}

\begin{minipage}{3.25in}{(c)\hspace*{5mm}\vspace*{-1.7in}
\psfig{file=Figure3cd.ps,width=3in,angle=0,clip=,bbllx=0pt,bblly=0pt,bburx=272pt,bbury=418pt}}\end{minipage} 
\begin{minipage}{3.25in}{\phantom{(d)}\hspace*{5mm}\vspace*{-1.7in}
\hspace*{3in}\phantom{a}}
\end{minipage} 

\vspace*{7mm}

\begin{minipage}{\textwidth}{
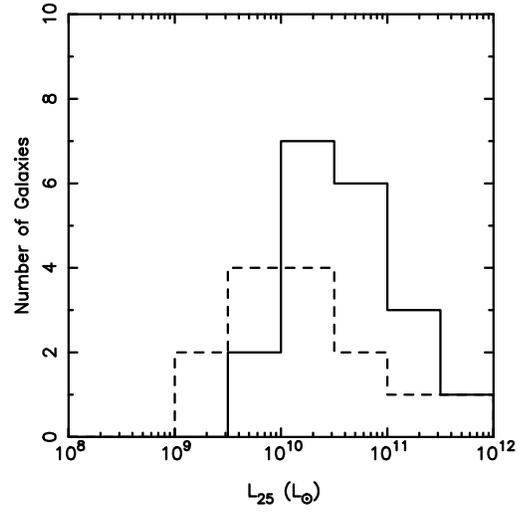
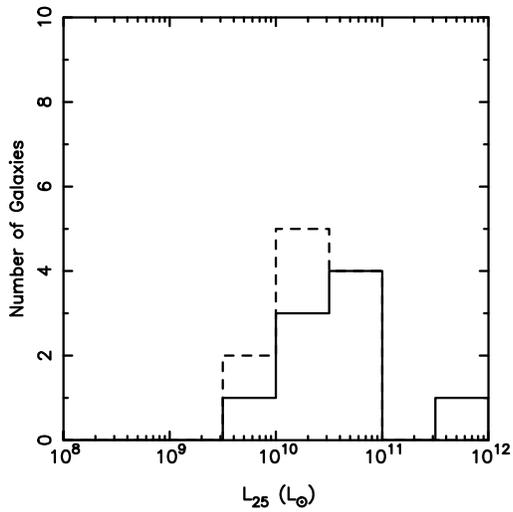
{\bf Figure 3:}  Luminosity distribution of the galaxies in which HBLRs are
detected (solid line) and not detected (dashed line) as a function of the
L$_{25}$ for the (a) Moran et al.\ (2000) sample, (b) Tran (2001)
sample and (c) the L01 sample.  
}\end{minipage}
\end{center}

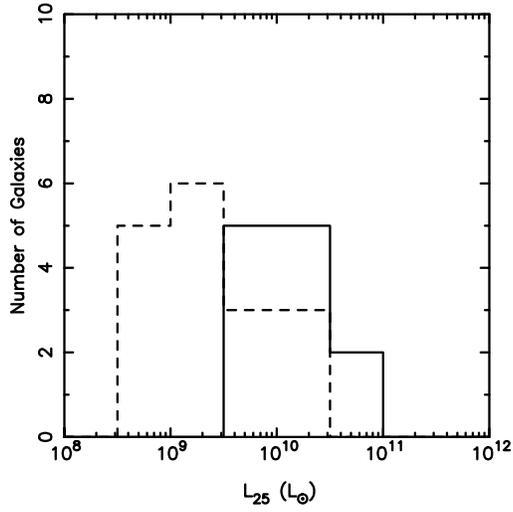
\end{document}